\documentclass[usenatbib]{mn2e}
\usepackage{times}
\usepackage{graphicx}
\usepackage{amsmath}
\usepackage{mathabx}
\usepackage{subfigure}
\usepackage{natbib}
\usepackage[colorlinks=true, linkcolor=blue, citecolor=blue, urlcolor=blue]{hyperref}
\usepackage{txfonts}

\newcommand{\aj}{AJ}
\newcommand{\mnras}{MNRAS}
\newcommand{\apj}{ApJ}
\newcommand{\apjs}{ApJS}
\newcommand{\apjl}{ApJL}
\newcommand{\aap}{A\&A}
\newcommand{\aaps}{A\&AS}

\newcommand{\nat}{Nature}
\newcommand{\pasp}{PASP}

\newcommand{\araa}{ARA\&A}

\defcitealias{2014MNRAS.441..869D}{DG14}

\begin{document} 

\title[Dust energy balance]{Dust energy balance study of two edge-on spiral galaxies in the Herschel-ATLAS survey}

\author[G. De Geyter et al.]
{%
Gert De Geyter,$^1$ 
Maarten Baes,$^1$ 
Ilse De Looze,$^{1,2}$
George J. Bendo,$^3$
Nathan Bourne,$^4$
\newauthor
Peter Camps,$^1$
Asantha Cooray,$^5$
Gianfranco De Zotti,$^{6,7}$
Loretta Dunne,$^{4,8}$,
\newauthor
Simon Dye,$^9$
Steve A. Eales,$^{10}$,
Jacopo Fritz,$^{1,11}$ 
Cristina Furlanetto,$^{9,12}$
\newauthor
Gianfranco Gentile,$^{1,13}$
Thomas M. Hughes,$^1$ 
Rob J. Ivison,$^{4,14}$
Steve J. Maddox,$^{4,8}$
\newauthor
Micha{\l}\ J. Micha{\l}owski,$^4$
Matthew W. L. Smith,$^{10}$
Elisabetta Valiante,$^{10}$
S\'ebastien Viaene$^1$
\\%
$^1$Sterrenkundig Observatorium, Universiteit Gent, Krijgslaan 281-S9, B-9000 Gent, Belgium \\
$^2$Institute of Astronomy, University of Cambridge, Madingley Road, Cambridge, CB3 0HA \\
$^3$UK ALMA Regional Centre Node, Jodrell Bank Centre for Astrophysics,
University of Manchester, Oxford Road, Manchester M13 9PL \\
$^4$Institute for Astronomy, University of Edinburgh, Royal Observatory, Blackford Hill, Edinburgh EH9 3HJ \\
$^5$Department of Physics and Astronomy, University of California, Irvine, CA 92697, USA\\
$^6$INAF-Osservatorio Astronomico di Padova, Vicolo Osservatorio 5, I-35122 Padova, Italy \\
$^7$SISSA, Via Bonomea 265, I-34136 Trieste, Italy \\
$^8$Department of Physics and Astronomy, University of Canterbury, Private Bag 4800, Christchurch, 8140, New Zealand\\
$^9$School of Physics and Astronomy, University of Nottingham, University Park, Nottingham NG7 2RD \\
$^{10}$School of Physics and Astronomy, Cardiff University, Queens Buildings, The Parade, Cardiff CF24 3AA \\
$^{11}$Centro de Radioastronom{\'\i}a y Astrof{\'\i}sica, UNAM Campus Morelia, Michoac\'an, Mexico \\
$^{12}$CAPES Foundation, Ministry of Education of Brazil, Brasilia/DF, 70040-020, Brazil \\
$^{13}$Department of Physics and Astrophysics, Vrije Universiteit Brussel, Pleinlaan 2, 1050 Brussels, Belgium\\
$^{14}$European Southern Observatory, Karl-Schwarzschild-Strasse 2, 85748 Garching bei M\"unchen, Germany
}

\date{\today}
\maketitle

\begin{abstract}
Interstellar dust in galaxies can be traced either through its extinction effects on the star light, or through its thermal emission at infrared wavelengths. Recent radiative transfer studies of several nearby edge-on galaxies have found an apparent inconsistency in the dust energy balance: the radiative transfer models that successfully explain the optical extinction underestimate the observed fluxes by an average factor of three. We investigate the dust energy balance for IC\,4225 and NGC\,5166, two edge-on spiral galaxies observed by the {\it{Herschel Space Observatory}} in the frame of the H-ATLAS survey. We start from models which were constrained from optical data and extend them to construct the entire spectral energy distribution of our galaxies. These predicted values are subsequently compared to the observed far-infrared fluxes. We find that including a young stellar population in the modelling is necessary as it plays a non-negligible part in the heating of the dust grains. While the modelling approach for both galaxies is nearly identical, we find two very different results. As is often seen in other edge-on spiral galaxies, the far-infrared emission of our radiative transfer model of IC\,4225 underestimates the observed fluxes by a factor of about three. For NGC\,5166 on the other hand, we find that both the predicted spectral energy distribution as well as the simulated images match the observations particularly well. We explore possible reasons for this difference and conclude that it is unlikely that one single mechanism is the cause of the dust energy balance problem in spiral galaxies. We discuss the different approaches that can be considered in order to get a conclusive answer on the origin this discrepancy.
\end{abstract} 


\section{Introduction}

Although interstellar dust accounts for only a small fraction of the interstellar medium (ISM), it plays an important role in various chemical and physical processes in galaxies. Apart from scattering and absorbing the light at ultraviolet (UV), optical and near-infrared (NIR) wavelengths, dust acts as a regulator of the physical and chemical conditions in the ISM. It is thus important to understand the amount and distribution of interstellar dust in galaxies.

When spiral galaxies are seen edge-on, their dust content often shows as a distinctive narrow lane where almost all of the stellar light is attenuated. Edge-on galaxies allow us to observe both the radial and vertical extent of the dust and stellar disk. Because the extinction effects are so clearly shown in edge-on spiral galaxies, they are the ideal targets when one wants to determine the dust content and distribution.

Studying the dust properties in extinction from UV, optical or NIR data offers some clear advantages.
The extinction properties of interstellar dust in this wavelength region are relatively well determined and, contrary to the far-infrared (FIR), data with excellent spatial resolution are easily obtained. The downside of this approach is the difficulty to determine the dust mass, as the complex star--dust geometry and physical processes like absorption or multiple anisotropic scattering make the conversion from attenuation to dust mass challenging \citep{1989MNRAS.239..939D, 1992ApJ...393..611W, 1994ApJ...432..114B, 1995MNRAS.277.1279D, 2001MNRAS.326..733B, 2005MNRAS.359..171I}. In cases where the contrasting light source is not the galaxy itself but a known background source, this conversion is more straightforward \citep{1992Natur.359..655W, 2009AJ....137.3000H, 2013MNRAS.433...47H}. In the last decade, much progress has been made in the field of 3D dust radiative transfer and a number of codes are now able to effectively incorporate the necessary physics \citep[e.g.,][]{2001ApJ...551..269G, 2003MNRAS.343.1081B, 2006MNRAS.372....2J, 2008A&A...490..461B, 2011A&A...536A..79R, 2012ApJ...746...70S}. An overview of the recent developments in 3D dust radiative transfer can be found in \cite{2013ARA&A..51...63S}.

Another way to determine the total dust mass is by measuring the thermal emission of the dust grains at mid-infrared (MIR), FIR and submm wavelengths. While there is still some uncertainty about the emissivity of dust grains at FIR/submm wavelengths,
global dust masses can be estimated either using simple modified blackbody fits 
or more advanced SED modelling tools. 
However, in this case, observations are less straightforward and the poor spatial resolution limits detailed information on the dust distribution to Local Group galaxies \citep{2010A&A...518L..71M, 2012A&A...543A..74X, 2012A&A...546A..34F, 2012MNRAS.423.2359D, 2012MNRAS.427..703S} or other nearby galaxies such as these from the Very Nearby Galaxy Survey \citep{2010A&A...518L..65B, 2012MNRAS.422.2291P, 2012MNRAS.419.1833B, 2012MNRAS.421.2917F, 2014A&A...565A...4H}.

As the starlight absorbed by dust grains in the UV/optical/NIR region is re-emitted in the MIR/FIR/submm wavelength regime, one would expect the absorbed luminosity to be equal to the total thermal IR/submm luminosity. However, there seems to be a deficit in this dust energy balance: the FIR emission predicted from extinction in edge-on spiral galaxies seems to underestimate the observed value by a factor of about three \citep{2000A&A...362..138P, 2011A&A...527A.109P, 2001A&A...372..775M, 2004A&A...425..109A, 2005A&A...437..447D, 2008A&A...490..461B, 2010A&A...518L..39B, 2012A&A...541L...5H}. This is a problem typically encountered using self-consistent radiative transfer simulations. SED fitting tools like MAGPHYS \citep{2008MNRAS.388.1595D} or CIGALE \citep{2005MNRAS.360.1413B, 2009A&A...507.1793N} do not encounter this problem. However, these methods cannot take into account the effects of the complex star-dust geometry in galaxies, which is necessary for a self-consistent calculation of the dust heating and thermal emission. Radiative transfer codes, on the other hand, can incorporate all the geometrical effects, position and strength of the heating sources to self-consistently predict the FIR emission. As this requires a significant increase in computational cost and with the loss of a guaranteed energy balance, they are less useful to study larger samples. While methods to predict results of radiative transfer simulations by using artificial neural networks are extremely promising, it is advisable to use them for statistical purposes \citep{2012MNRAS.423..746S}.
 
One possible explanation for the discrepancy in the energy balance found in radiative transfer simulations is a systematic underestimation of the FIR emission coefficients \citep{2004A&A...425..109A, 2005A&A...437..447D, 2011ApJ...741....6M}. Other authors have advocated the presence of additional dust distributed in such a way that it hardly contributes to the attenuation, either in the form of a second inner disc \citep{2000A&A...362..138P, 2011A&A...527A.109P, 2001A&A...372..775M}, or as a clumpy or inhomogeneous medium \citep{2008A&A...490..461B, 2012MNRAS.427.2797D, 2014A&A...571A..69D, 2015A&A...576A..31S}. Finally, there is now ample evidence for the presence of vertically extended diffuse dust in spiral galaxies, which can contribute at least some of the infrared emission without contributing much to the large-scale attenuation \citep{2007A&A...471L...1K, 2009MNRAS.395...97W, 2014ApJ...785L..18S, 2014ApJ...789..131H}.

In order to explain this discrepancy, we require a sufficiently large set of galaxies that have the needed multi-wavelength data available and are modelled in a homogeneous way, such that physical effects can be disentangled from effects connected to the modelling. Since fitting radiative transfer models to images is a very computationally expensive job, most dust energy balance studies have concentrated on a single or at most a very modest set of galaxies. Moreover, the parameter space of the different galaxy models is often degenerate, in the sense that strongly different galaxy models can result in similar attenuation maps \citep{2007A&A...471..765B, 2013A&A...550A..74D}.  Fortunately, these dust energy balance studies can now be taken to the next level, thanks to two recent developments.

The first improvement concerns the determination of the dust mass based on optical data. Recently, different attempts have been undertaken in the radiative transfer community to make the fitting of radiative transfer models to data more automated and objective \citep[e.g.,][]{2005A&A...434..167S, 2007ApJS..169..328R, 2007A&A...471..765B, 2012ApJ...746...70S}. One of the challenges here is that, in order to fit Monte Carlo radiative transfer models to data, one has to take into account that the model images themselves are not noise-free, which implies that normal gradient-search optimisation methods are problematic. \citet{2013A&A...550A..74D} showed that optimisation by means of genetic algorithms is an ideal solution for this problem. Moreover, the degeneracy in the parameter space can be minimised by means of  oligochromatic radiative transfer modelling, i.e., by simultaneously fitting radiative transfer models to images in several optical wavebands \citep[][hereafter \citetalias{2014MNRAS.441..869D}]{2014MNRAS.441..869D}. 

The second development is the launch of the {\it{Herschel Space Observatory}} \citep{2010A&A...518L...1P}, which has been a major boost for the study of the interstellar dust medium in nearby galaxies. Particularly revolutionary is the SPIRE photometer, which filled a gap in the wavelength coverage between previous FIR missions as IRAS \citep{1984ApJ...278L...1N}, ISO \citep{1996A&A...315L..27K} and Spitzer \citep{2004ApJS..154....1W} on the one side, and ground-based submm instruments as SCUBA \citep{1999MNRAS.303..659H} and LABOCA \citep{2009A&A...497..945S} on the other side. Thanks to {\it{Herschel}}, we can finally construct complete dust SEDs, easily trace even the cold dust grains and get a more realistic and tighter constraint on the total dust mass. 

The target galaxies in this paper were observed as part of the H-ATLAS survey \citep{2010PASP..122..499E}. This was the largest extragalactic survey undertaken by {\it{Herschel}}, with a total of 600 hours dedicated observing time. In total, about 600~deg$^2$ of extragalactic sky was mapped simultaneously with the PACS \citep{2010A&A...518L...2P} and SPIRE \citep{2010A&A...518L...3G} instruments in 5 bands centred at 100, 160, 250, 350 and 500 $\mu$m. The scientific goals of the H-ATLAS programme are diverse, ranging from the search for debris disks and the characterisation of the Galactic cirrus \citep{2010A&A...518L.134T, 2011MNRAS.412.1151B}, statistical studies of the local and the high-redshift galaxy population \citep{2011MNRAS.417.1510D, 2011ApJ...742...24L, 2012MNRAS.427..703S}, to the investigation of extreme objects like blazars and gravitational lenses \citep{2010Sci...330..800N, 2011ApJ...740...63C, 2012ApJ...753..134F, 2013MNRAS.430.1566L}. 

In this paper, we will make use of these two recent developments to investigate the energy balance in two edge-on spiral galaxies, IC\,4225 and NGC\,5166. In Section~{\ref{obs.sec}} we describe the rationale behind the selection of these two galaxies, and present the available data. The construction and results of our radiative transfer modelling is presented in Section~{\ref{meth.sec}}. In Section~{\ref{disc.sec}} we discuss these results and the possible implications on the dust energy balance problem and present our conclusions.

\section{The targets}
\label{obs.sec}

\subsection{Target selection}

\begin{table}
\centering
\begin{tabular}{cccc}
\hline\hline\\[-1ex]
Property & unit & IC\,4225 & NGC\,5166 \\[1ex]
\hline \\[-2ex]
$\alpha_{2000}$ & h & 13:20:01 & 13:28:15 \\
$\delta_{2000}$ & deg & 31:58:53 & 32:01:57 \\
type & --- & S0/a & Sb\\
$v_{\text{r}}$ & km\,s$^{-1}$ & 5438 & 4647 \\ 
$d$ & Mpc & 73.9 & 62.6 \\
scale & kpc\,arcsec$^{-1}$ & 0.358 & 0.303 \\
$D_{25}$ & arcmin & $1.3\times0.3$ & $2.3\times0.4$ \\
$\log L_{\text{K}}$ & $L_{\odot,\text{K}}$ & 10.93 & 11.19 \\
$\log M_\star$ & $M_\odot$ & 10.59 & 10.82 \\
$\log L_{\text{TIR}}$ & $L_\odot$ & 9.74 & 10.30 \\
{{$\log M_{\text{d}}$}} & {{$M_\odot$}} & {{7.24}} & {{7.89}} \\
$\log {\text{SFR}}$ & $M_\odot$\,yr$^{-1}$ & --0.58 & --0.03 \\
$\log {\text{sSFR}}$ & Gyr$^{-1}$ & --2.17 & --1.87 \\
\hline\hline
\end{tabular}
\caption{
The fundamental properties of the two galaxies in our sample. Stellar masses, TIR luminosities, {{dust masses,}} star formation rates and specific star formation rates are calculated using a MAGPHYS modelling of the spectral energy distribution.
}
\label{Properties.tab}
\end{table}

\begin{figure}
\centering
\includegraphics[width=\columnwidth]{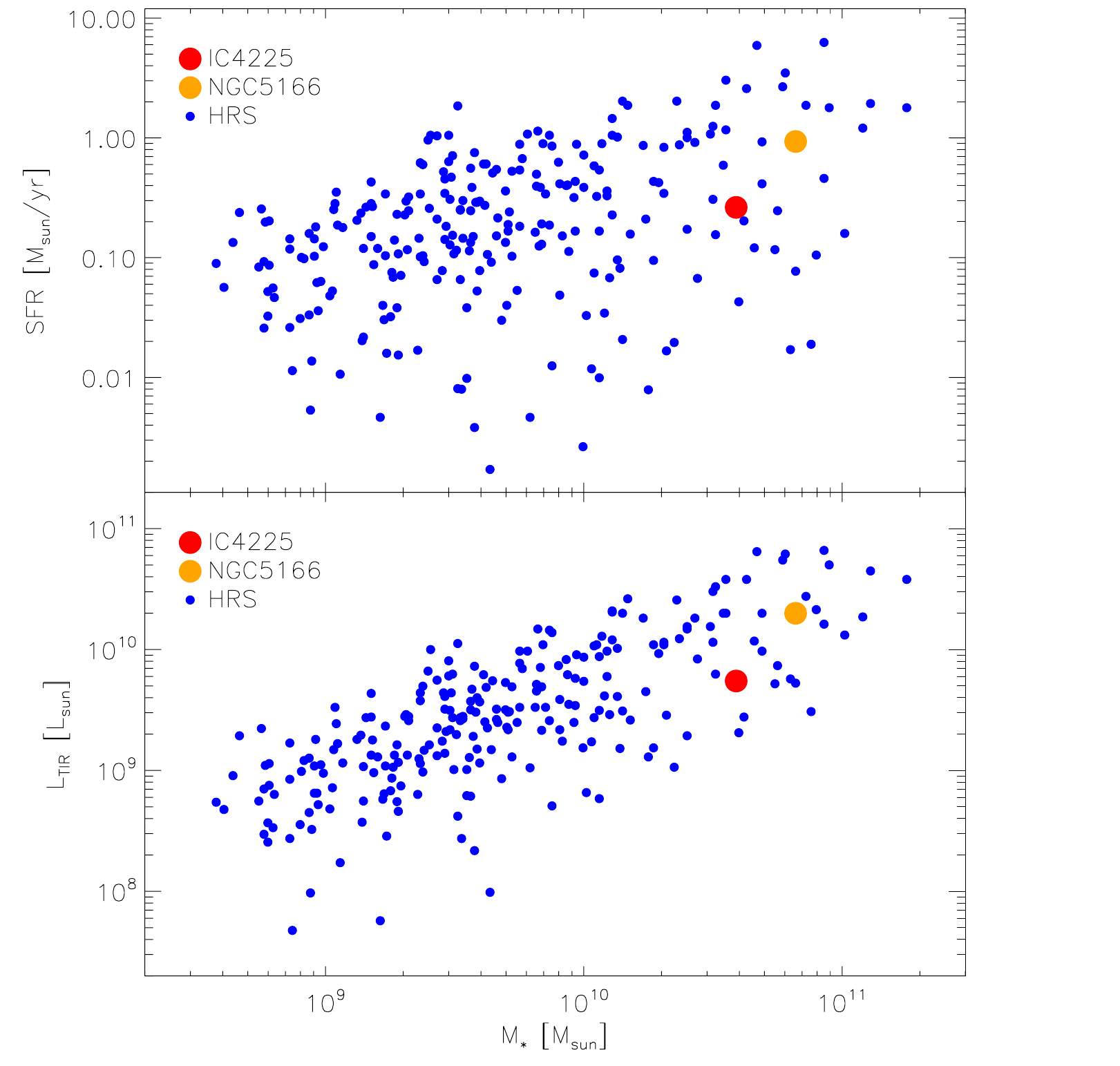}
\caption{
A comparison of some fundamental properties of the two galaxies in our sample and the HRS late-type galaxy population. The top panel compares stellar mass and star formation rate, the bottom panel stellar mass and dust luminosity. All quantities have been computed in a similar way using MAGPHYS.
}
\label{CompareHRS.fig}
\end{figure}

The starting point for our target selection was the sample of 12 edge-on spiral galaxies considered by \citetalias{2014MNRAS.441..869D}. This sample was selected from the CALIFA survey \citep{2012A&A...538A...8S}, based on the criteria of an edge-on orientation with a dust-lane morphology, a major axis diameter larger than 1~arcmin and a minor axis diameter larger than 8~arcsec. Each of these galaxies were modelled in exactly the same way: a stars+dust radiative transfer model was fitted to the SDSS {\em{g}}, {\em{r}}, {\em{i}} and {\em{z}} images, using the oligochromatic FitSKIRT code. 

In this paper, we want to extend this analysis to a panchromatic study, and compare the dust ``observed'' in the optical SDSS images to the dust seen in emission in the FIR/submm. We hence select from the sample of \citetalias{2014MNRAS.441..869D} those galaxies that have {\it{Herschel}} data to cover the FIR/submm SED. 

Of the 12 galaxies, three galaxies have been observed by {\em{Herschel}}: IC\,2461, IC\,4225, and NGC\,5166. We decided to leave out IC\,2461 from this study -- it turned out to be an extreme system with a deeply buried Seyfert~2 nucleus \citep{2012A&A...545A.101P}, that is emitting high-energy emission even in the ultra-hard X-ray bands \citep{2011ApJ...739...57K, 2013ApJS..207...19B}. Additionally, we lack the required UV observations from GALEX, which provides an important constraint on the younger stellar population. 

Both of the remaining galaxies are located in the North Galactic Plane (NGP) field of the H-ATLAS survey. IC\,4225 is situated in the Coma supercluster region at a distance of about 73.9 Mpc \citep{1988A&AS...74...83S,1995AJ....109.1458R}. It is classified as an S0/a galaxy \citep{1991rc3..book.....D} and is about 1.3  by 0.3 arcmin. NGC\,5166 is an Sb type galaxy at a Tully-Fisher based distance of 62.6 Mpc \citep{1991rc3..book.....D, 2007A&A...465...71T}. It has a prominent but somewhat asymmetric dust lane and has a larger angular extent of 2.3 by 0.4 arcmin. Table~{\ref{Properties.tab}} lists a number of fundamental properties of the two galaxies. Stellar masses, dust luminosities, star formation rates and specific star formation rates are calculated by fitting the UV-submm spectral energy distribution using MAGPHYS \citep{2008MNRAS.388.1595D}. In Figure~{\ref{CompareHRS.fig}} we plot the location of the two galaxies in the $M_*$--SFR and the $M_*$--$L_{\text{TIR}}$ planes. Overplotted on these panels is the location of the late-type galaxy population from the {\it{Herschel}} Reference Survey \citep[HRS,][]{2010PASP..122..261B}, a complete volume-limited survey of galaxies in the local Universe. The plotted quantities are calculated in the same way (Viaene et al., {\it{in prep.}}). Table~{\ref{Properties.tab}} and Figure~{\ref{CompareHRS.fig}} show that IC\,4225 and NGC\,5166 are relatively massive spiral galaxies with a high dust content and star-formation rate. This is not unexpected, given that one of the selection criteria of the parent sample from \citetalias{2014MNRAS.441..869D} was the presence of a clear dust lane. 

\subsection{Data}

\begin{table}
\centering
\begin{tabular}{ccr@{$\,\pm\,$}lr@{$\,\pm\,$}lr@{$\,\pm\,$}lr@{$\,\pm\,$}lr@{$\,\pm\,$}lr@{$\,\pm\,$}l}
\hline\hline\\[-1ex]
Survey & wavelength & 
\multicolumn{2}{c}{IC\,4225} &
\multicolumn{2}{c}{NGC\,5166}\\
& [$\mu$m] & 
\multicolumn{2}{c}{[mJy]} &
\multicolumn{2}{c}{[mJy]}\\[1ex]
\hline \\[-2ex]
GALEX&0.226& 0.40 & 0.12 & 0.87 & 0.17  \\
SDSS&0.354& 1.3 & 0.1 & 2.4 & 0.1\\
SDSS&0.475& 5.73 & 0.1 & 11.9 & 0.2 \\
SDSS&0.622& 12.3 & 0.2 & 26.1 & 0.3 \\
SDSS&0.763& 19.0 & 0.3 & 40.5 & 0.4\\
SDSS&0.905& 26.6 & 0.3 & 55.0 & 0.4\\
2MASS&1.25& 40.3 & 0.5 & 80.0 & 0.7 \\
2MASS&1.64& 53.0 & 0.5 & 127.6 & 1.0 \\
2MASS&2.17& 47.0 & 0.5 & 119.6 & 0.9 \\
WISE &3.4& 23.0 & 0.2 & 51.1 & 0.3 \\
WISE &4.6& 13.6 & 0.2 & 33.0 & 0.3\\
WISE &12& 15.5 & 0.2 & 85.8 & 0.4\\
WISE &22& 11.5 & 0.8 & 170 & 3.0 \\
IRAS &25& \multicolumn{2}{c}{---} & 155 & 38 \\
IRAS &60& \multicolumn{2}{c}{---} & 845 & 59 \\
IRAS &100& \multicolumn{2}{c}{---} & 2880 & 259 \\
PACS &100& 640 & 72 & 3132 & 252 \\
PACS &160& 709 & 85 & 3506 & 262 \\
SPIRE &250& 490 & 35 & 2315 & 167 \\
SPIRE &350& 190 & 15 & 961 & 74 \\
Planck &350& \multicolumn{2}{c}{---} & 1062 & 272 \\
SPIRE &500& 54 & 7 & 327 & 33 \\
Planck &550& \multicolumn{2}{c}{---} & 356 & 219 \\[1ex]
\hline\hline
 \end{tabular}
 \caption{Observed flux densities and corresponding errors for IC\,4225 and NGC\,5166.}
\label{Fluxes.tab}
\end{table}

For the panchromatic radiative transfer modelling that we will perform, we need imaging data over a wavelength range as broad as possible. Both galaxies have been observed by the Sloan Digital Sky Survey \citep[SDSS,][]{2009ApJS..182..543A} and therefore have $u$, $g$, $r$, $i$ and $z$ band observations. Additionally they were observed by the Galaxy Evolution Explorer \citep[GALEX,][]{2005ApJ...619L...1M}, the Two Micron All Sky Survey \citep[2MASS,][]{2006AJ....131.1163S} and the Wide-field Infrared Survey Explorer \citep[WISE,][]{2010AJ....140.1868W}. Unfortunately, neither of the galaxies had a clear detection in the GALEX FUV band at 0.152~$\mu$m, so only the NUV flux density of the GALEX survey could be used. For each of these observations, we followed standard data reduction procedures using IDL routines. After masking foreground stars, integrated flux densities were determined by means of aperture photometry using DS9 Funtools \citep{2001ASPC..238..225M} instead of using the catalog values to ensure that there is no contamination from nearby sources. 

PACS and SPIRE images and flux densities were extracted from the H-ATLAS catalogue. Details on the PACS and SPIRE data reduction, map-making, source extraction and catalogue generation for H-ATLAS can be found in \citet{2010MNRAS.409...38I}, \citet{2011MNRAS.415..911P}, \citet{2011MNRAS.415.2336R} and Valiante et al.\ ({\it{in prep.}}). For NGC\,5166 additional data were used from the IRAS Faint Source Catalog \citep{1992ifss.book.....M} at 25, 60 and 100 $\mu$m. The galaxy was also detected at 350 and 550 $\mu$m by {\em{Planck}} where we have used the flux density determined using the aperture photometry (APERFLUX) listed in the Planck Catalogue of Compact Sources v1.0 \citep{2014A&A...571A..28P}.  

All the flux densities and their corresponding errors for IC\,4225 and NGC\,5166 are listed Table \ref{Fluxes.tab}.

\section{Panchromatic radiative transfer modelling}
\label{meth.sec}

\subsection{Modelling approach}

\citetalias{2014MNRAS.441..869D} determined the dust distribution in 12 edge-on spiral galaxies, including IC\,4225 and NGC\,5166, based on optical images. More specifically, they used the oligochromatic FitSKIRT code to construct a radiative transfer model with 19 free parameters that can simultaneously reproduce the SDSS {\em{g}}, {\em{r}}, {\em{i}} and {\em{z}} band data for these galaxies. For more details on the fitting procedure we refer to Section 3 of \citetalias{2014MNRAS.441..869D}. The goal of the modelling in the present paper is to extend these oligochromatic models to fully panchromatic models that can reproduce not only the images at optical wavelengths, but also explain the thermal emission by dust, as observed in the MIR/FIR/submm in a self-consistent way. 

The modelling is done using SKIRT \citep{2003MNRAS.343.1081B, 2011ApJS..196...22B, 2015A&C.....9...20C}, a Monte Carlo radiative transfer code designed to solve the 3D dust radiative transfer problem.  
It is mainly applied to simulate dusty galaxies \citep[e.g.,][]{2010MNRAS.403.2053G, 2010A&A...518L..45G, 2012MNRAS.419..895D, 2014A&A...571A..69D}, but has also been used to model active galactic nuclei \citep{2012MNRAS.420.2756S} and dusty discs around evolved stars \citep{2007BaltA..16..101V, 2015A&A...577A..55D}. 

Our modelling goes in three phases: our first set of models (Case A: Optical-NIR models, Section~{\ref{CaseA.sec}}) are constrained by the optical images only. In a second set of models (Case B: UV--NIR models, Section~{\ref{CaseB.sec}}), we add a component to account for additional UV radiation. Finally, we construct a third set of models where we add ad hoc components for the cool dust and obscured star formation in order to fit the excesses that are still present between the Case~B models and the observed infrared flux densities (Case C: UV--NIR ModBB models, Section~{\ref{CaseC.sec}}).

\begin{table}
\centering
\begin{tabular}{ccr@{$\,\pm\,$}lr@{$\,\pm\,$}l}
\hline\hline\\[-1ex]
Parameter & unit & 
\multicolumn{2}{c}{IC\,4225} & 
\multicolumn{2}{c}{NGC\,5166} \\[1ex]
\hline \\[-2ex]
$h_{R,*,{\text{ev}}}$ & kpc & 3.38 & 0.05 & 3.92 & 0.17\\
$h_{z,*,{\text{ev}}}$ & kpc &  0.84 & 0.04 & 0.66 & 0.04\\
$R_{\text{eff}}$ & kpc & 1.25 & 0.04 & 3.17 & 0.59\\
$n$ & --- & 1.1 &0.1 & 5.8 & 0.9\\
$q$ & --- & 0.69 & 0.03 & 0.26 & 0.04\\
$h_{R,\text{d}}$ & kpc & 10.01 & 2.74 & 5.81 & 1.12\\
$h_{z,\text{d}}$ & kpc & 0.24 & 0.02 & 0.34 & 0.06\\
$M_{\text{d}}$ & $10^7~M_\odot$ & 2.1 & 0.5 & 4.8 & 0.9\\
$\tau_{\text{V}}^{\text{f}}$ & --- & 0.18 & 0.17 & 1.23 & 0.37\\
$\tau_{\text{V}}^{\text{e}}$ & --- & 7.4 & 0.8 & 21.2 & 4.7\\
$i$ & deg & 89.3 & 0.1 & 87.6 & 0.5\\
$L_{i}^{\text{tot}}$ & $10^9~L_{\odot}$ & 4.16 & 0.1 & 11.39 & 0.51\\ 
$B/T_{i}$ & --- & 0.43 & 0.01 & 0.45 & 0.02\\
${\text{SSP}}_{{\text{disk}}}$ & Gyr & \multicolumn{2}{c}{7} & \multicolumn{2}{c}{4} \\
${\text{SSP}}_{{\text{bulge}}}$ & Gyr & \multicolumn{2}{c}{12} & \multicolumn{2}{c}{11} \\
$h_{R,*,{\text{young}}}$ & kpc & \multicolumn{2}{c}{10.01} & \multicolumn{2}{c}{5.81} \\
$h_{z,*,{\text{young}}}$ & kpc & \multicolumn{2}{c}{0.12} & \multicolumn{2}{c}{0.17} \\
$T_{\text{d,ModBB}}$ & K & \multicolumn{2}{c}{21} & \multicolumn{2}{c}{20} \\
$M_{\text{d,ModBB}}$ & $10^7~M_\odot$ & \multicolumn{2}{c}{0.9} & \multicolumn{2}{c}{1.5} \\[1ex]
 \hline\hline
 \end{tabular}
 \caption{The parameters of the best fitting models for IC\,4225 and NGC\,5166. The top 13 parameters with error bars correspond to the FitSKIRT models as described in \citetalias{2014MNRAS.441..869D}. The bottom six parameters are the additional parameters introduced for the panchromatic radiative transfer models: the SSP ages for disk and bulge for the Case~A models, the scale length and scale height of the young stellar disc for the Case~B models, and the parameters of the modified black body dust emission for the Case~C models.}
\label{Param.tab}
\end{table}

\subsection{Case~A: Optical--NIR models}
\label{CaseA.sec}

\begin{figure*}
\centering
\includegraphics[width=0.75\textwidth]{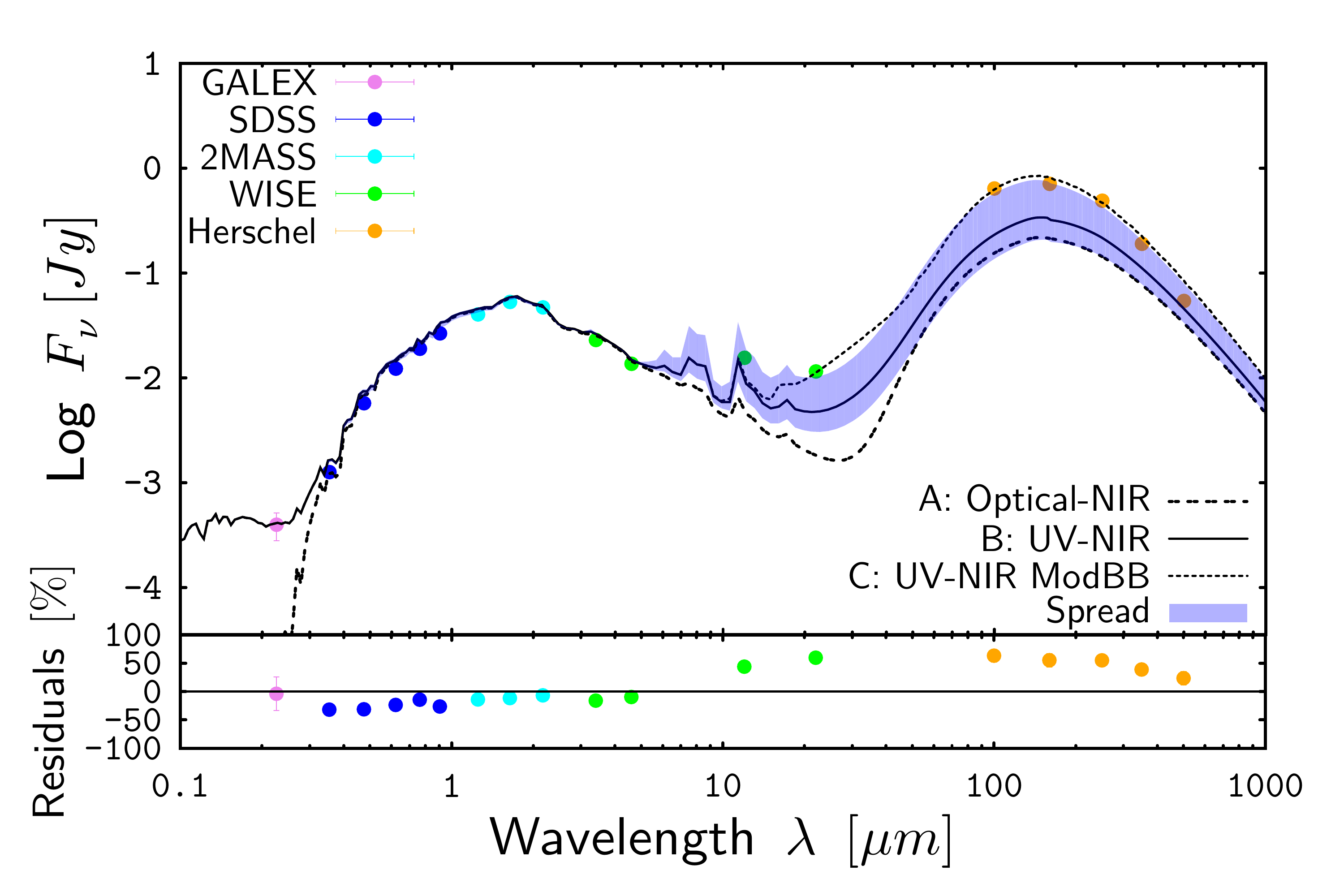}
\includegraphics[width=0.75\textwidth]{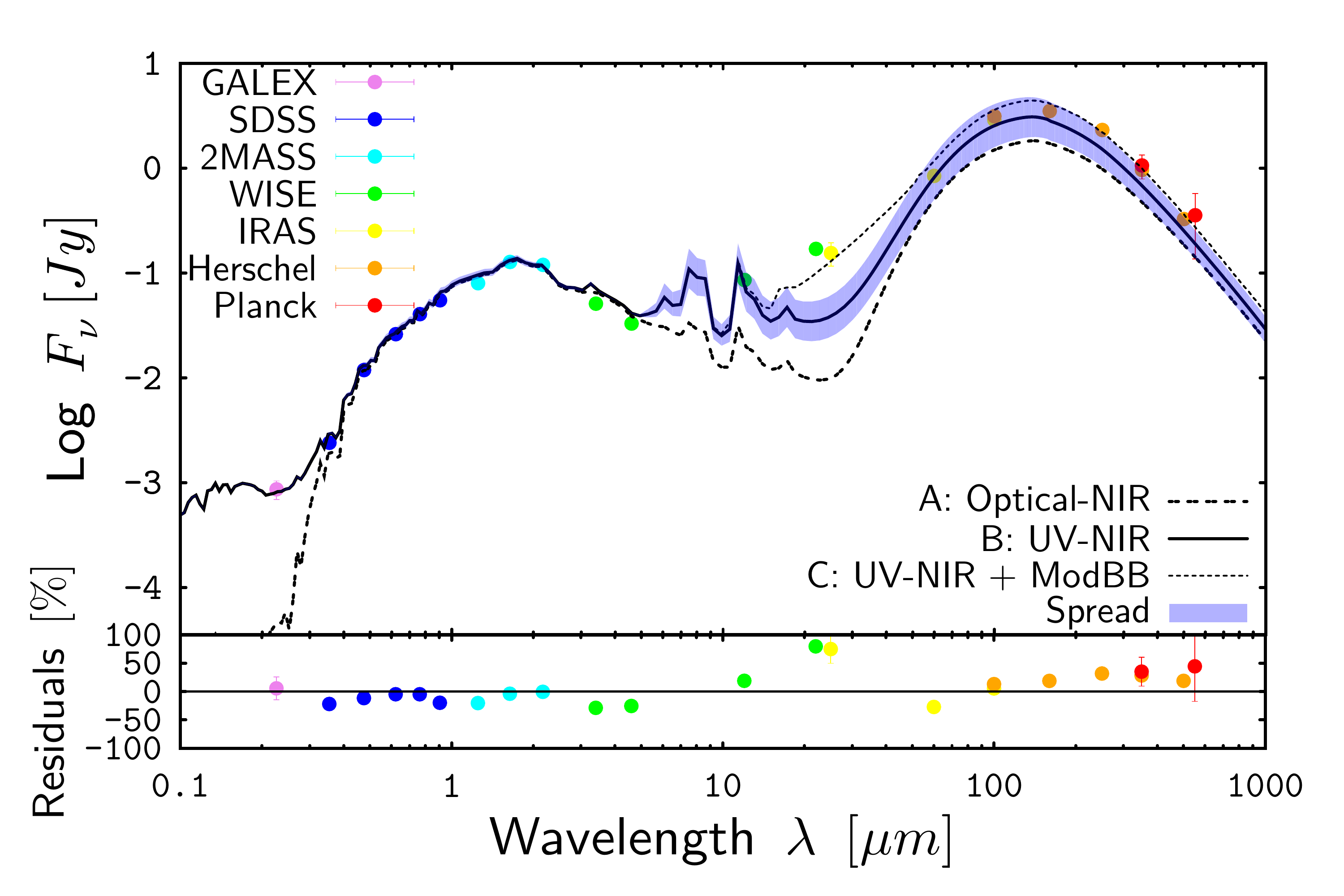}
\caption{The SED of IC\,4225 (top) and NGC\,5166 (bottom). The coloured dots with error bars correspond to the flux densities listed in Table~\ref{Fluxes.tab}. The {{dashed}} lines are the simple models based only on the optical data (Case~A), the solid lines are the models in which an additional young stellar component in taken into account (Case~B), and the {{dotted}} lines are the models with additional cold dust and obscured star formation (Case~C). The blue transparent band represent the spread in the FIR emission of the Case~B models within one error bar of the best fitting models. The bottom panels below the SEDs show the residuals between the observed SED and the Case~B model.}
\label{SEDs.fig}
\end{figure*}

The FitSKIRT results from \citetalias{2014MNRAS.441..869D} form the ansatz for the models. The stellar distribution consists of an double-exponential disk to describe the stellar disk and a flattened S\'ersic profile to describe the bulge. The dust distribution is represented by a double-exponential disk with scale length $h_{R,{\text{d}}}$ and scale height $h_{z,{\text{d}}}$. This is a commonly used distribution in dust radiative transfer simulations \citep{1999A&A...344..868X, 2001A&A...372..775M, 2004A&A...419..821T, 2007A&A...471..765B, 2011A&A...527A.109P}. Recently, it has been shown, using high resolution PACS and SPIRE maps, that such a distribution forms a suitable description of the FIR morphology in both radial and vertical direction \citep{2013A&A...556A..54V, 2014A&A...565A...4H}. The standard BARE-GR-S model of \citet{2004ApJS..152..211Z} was used to set the absorption efficiency, the scattering efficiency and scattering phase function of the dust. The total dust mass $M_{\text{d}}$ completes the characterisation of the dust content in the model. 

One remarkable result from this fitting is that IC\,4225 has an exceptionally large dust disk as can be seen in Table~\ref{Param.tab}. Especially when compared in relative size to the stellar disk, it is almost three times more extended. It is even more surprising keeping in mind that this galaxy was classified as a S0/a galaxy, for which it has been shown that the dust emission should be compact \citep{2007MNRAS.380.1313B, 2009ApJ...703.1569M, 2012ApJ...748..123S}. Most likely however, due to the smaller angular size, the morphology of IC\,4225 was misclassified. 

The large dust disk, combined with a total dust mass close to the average of the modelled sample, results in a face-on V-band optical depth of a mere 0.18. Because of the relatively small angular extent of the galaxy (it is the galaxy with the smallest angular extent of the sample considered by \citetalias{2014MNRAS.441..869D}), the uncertainties on the dust parameters are quite large compared to the average values for the sample. In spite of the large uncertainties, the model reproduces the optical images to an acceptable level with only a small irregularity in the outer regions, possibly due to the projection of a spiral structure (see Figure~3{\em{i}} of \citetalias{2014MNRAS.441..869D}).

The model for NGC\,5166 is more in line with those found for other edge-on spiral galaxies. The dust disk is about 50\% more extended and about half as thick as the stellar disk, which corresponds to the classical image of edge-on spiral galaxies \citep{1999A&A...344..868X, 2007A&A...471..765B}. The dust parameters were constrained with a higher accuracy compared to IC\,4225. The model has a bulge that is quite large, relative to the disk. As was found by \cite{2000A&A...362..435L} and can be clearly seen from the residuals in Figure 3{\em{j}} in \citetalias{2014MNRAS.441..869D}, the bulge has a strong peanut shape, a feature that is quite common when looking at edge-on spiral galaxies \citep{2000A&AS..145..405L}. As was the case for IC\,4225, a small asymmetry can be seen, possibly due to a spiral structure. In general, the fit is very satisfactory with most pixels only showing a maximum deviation of 25\%.

The first step in our analysis consists of adjusting these oligochromatic FitSKIRT models in order to calculate the view of the galaxy in the entire UV to submm domain, rather than only in a few selected optical bands. This implies that both the properties for the stars and dust need to be set over this entire wavelength domain. 

For both stellar components, we assume an intrinsic \citet{2003MNRAS.344.1000B} single stellar population SED, which is based on \citet{2003PASP..115..763C} initial mass function and assuming a solar metallicity. The ages of the best fitting SSPs were determined from the dereddened luminosities reported in \citetalias{2014MNRAS.441..869D} using an adapted version of the SED fitting tool from \citet{2008MNRAS.386.1252H, 2009MNRAS.399.1206H}. For IC\,4225, we found ages of 7~Gyr for the stellar disk and 12~Gyr for the bulge component. The best fitting stellar populations for NGC\,5166 are slightly younger at 4~Gyr for the stellar disk and 11~Gyr for the bulge component. The entire SED was normalised to match the {\em{i}}-band luminosity. 

Concerning the dust, the standard BARE-GR-S dust model was used to set the optical properties in the entire UV--submm wavelength domain. Note that the SKIRT code considers individual silicate, graphite and PAH grain populations for the calculation of the dust emission spectrum, and that it fully takes into account the transient heating of PAHs and very small grains, using techniques described in \citet{1989ApJ...345..230G} and \citet{2001ApJ...551..807D}. 

In Figure~{\ref{SEDs.fig}} we compare the SED obtained for the Case~A model ({{dashed}} lines) with the observed flux densities for both galaxies. It is clear that for both galaxies the models cannot explain the observed SEDs. The radiative transfer models reproduce the optical SED very well, which is not surprising given that the geometrical parameters of the model and the characteristics of the intrinsic SEDs were determined to reproduce the optical images. They can also successfully explain the NIR data points. The model SEDs, however, underestimate the observed flux densities both in the UV and in the thermal MIR/FIR/submm regions.

\subsection{Case~B: UV--NIR models}
\label{CaseB.sec}

It is clear that the simple Case~A models considered above need to be extended with an additional component that can account for the observed UV radiation. We will account for this by adding a third stellar component associated with young, unobscured star forming regions. Unfortunately, for both galaxies, the GALEX NUV images are too noisy to be used to constrain the geometry of the younger component without any prior assumptions. Young stellar populations in spiral galaxies are usually located in a thin disk with scale heights similar or smaller than the dust disk \citep{2013ApJ...773...45S, 2014ApJ...795..136S}. In our models, the scale height of this young stellar disk is taken to be half that of the dust disk\footnote{For both galaxies we ran additional simulations with different values for the scale height of the young stellar population, ranging between one third of the dust scale height to the same value. This effect, however, was hardly noticeable, as was also found by \citet{2014A&A...571A..69D}.}. The values we use are 120 and 170~pc for IC\,4225 and NGC\,5166 respectively, similar to what was found as the scale height of the young stellar disk in the Milky Way \citep{1980ApJS...44...73B}.

For the intrinsic SED of this young component, we used a Starburst99 SED template which represents a stellar population with a constant, continuous star formation rate and an evolution up to 100 Myr \citep{1999ApJS..123....3L}. In this case, the initial mass function is a \citet{1955ApJ...121..161S} IMF with masses between $1~M_\odot$ and $100~M_\odot$ and again assuming a solar metallicity. The luminosity of this component is constrained by the GALEX NUV flux density. As the contribution of this SED is negligible in the optical bands, adding this component does not affect the models in the SDSS and 2MASS bands. As it does, however, affect the energy balance of the dust, a new set of radiative transfer simulations has to be run in order to constrain the NUV flux and determine the resulting images over the entire wavelength range.

The solid line in the left panel of Figure~{\ref{SEDs.fig}} compares the model SED of this new Case~B model with the observed flux densities for IC\,4225. Compared to the simple model without young stellar component, this model is a clear improvement. The optical/NIR SED is still reproduced as accurately as in the simple model, but the model now also reproduces the UV portion of the SED. As a result of the additional energy input of the young stellar populations, much of which is absorbed by the dust, the infrared part of the SED is also boosted. However, it is clear that a FIR excess still remains: the model FIR/submm flux densities still underestimate the observed data points by a factor of a few. 

Interestingly, we obtain a different picture for NGC\,5166. There was also a clear FIR/submm inconsistency between the data and the simple Case~A model. Contrary to IC\,4225, however, this inconsistency disappears almost completely when we include a young stellar component: not only does the model now reproduce the UV part of the SED, but it almost provides sufficiently additional dust emission to account for the observed FIR/submm emission. The only data points that are still strongly underestimated by the model are the WISE 22~$\mu$m and IRAS 25~$\mu$m, which are usually linked to obscured star formation \citep{2007ApJ...666..870C, 2010ApJ...714.1256C, 2013ApJ...774...62L}. 

To investigate the effect of the uncertainty on the dust distribution of IC\,4225, we have modelled the spread in the FIR emission when starting from a model with slightly altered parameter values. Table 3 in \citetalias{2014MNRAS.441..869D} mentions both the best fitting values and the uncertainties for every free parameter. The highest FIR emission can be expected to be the model with the smallest dust disk and highest dust mass while the lowest FIR emission is found in the opposite case. Figure~{\ref{SEDs.fig}} shows the spread on the Case B model where the luminosity of the young stellar component was adjusted to reproduce the original value. By using this approach we investigate the FIR emission of models that still appropriately model the optical bands and make no additional assumptions using MIR/FIR data.

It is clear that for IC\,4225, the spread on the FIR emission is larger, resulting from the higher uncertainty on the dust distribution. The observed data points are only barely compatible with the densest dust disk model. However, in this case even the $22\ \mu m$ flux is recovered. For NGC\,5166, the spread in the FIR is considerably smaller and not much variation on the shape of the SED can be expected. 

\begin{figure*}
\centering
\hspace*{\fill}
\subfigure{\includegraphics[width=0.46\textwidth]{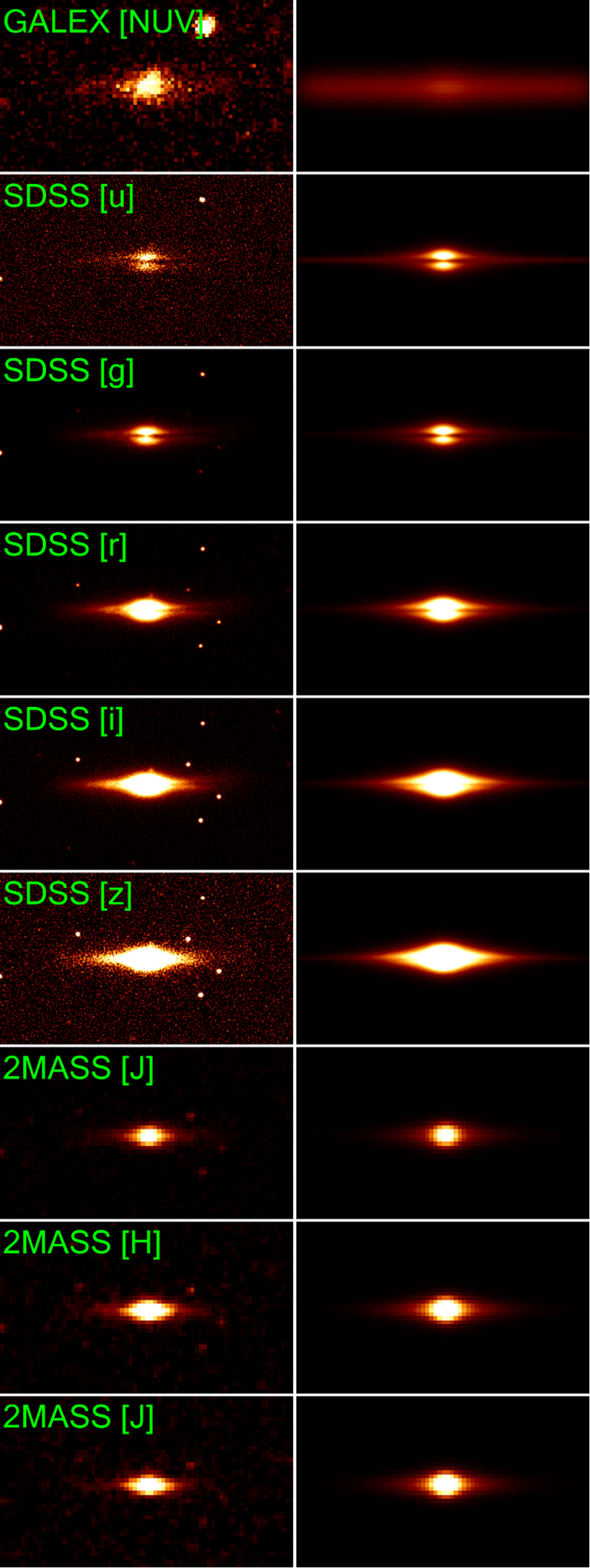}}\hfill
\subfigure{\includegraphics[width=0.46\textwidth]{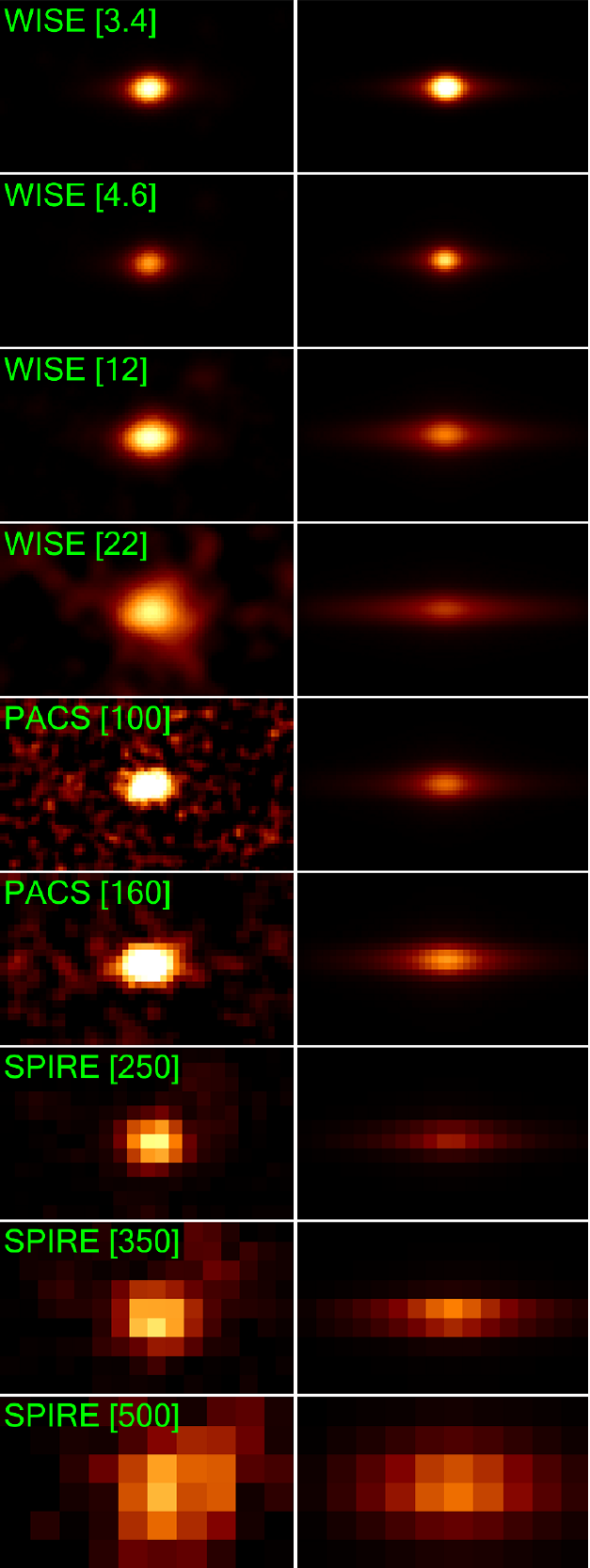}}
\hspace*{\fill}
\caption{Comparison between the observations (left) and model images (right) for IC\,4225. All images correspond to the Case~B model.}
\label{comparison_IC4225.fig}
\end{figure*}

\begin{figure*}
\centering
\hspace*{\fill}
\subfigure{\includegraphics[width=0.46\textwidth]{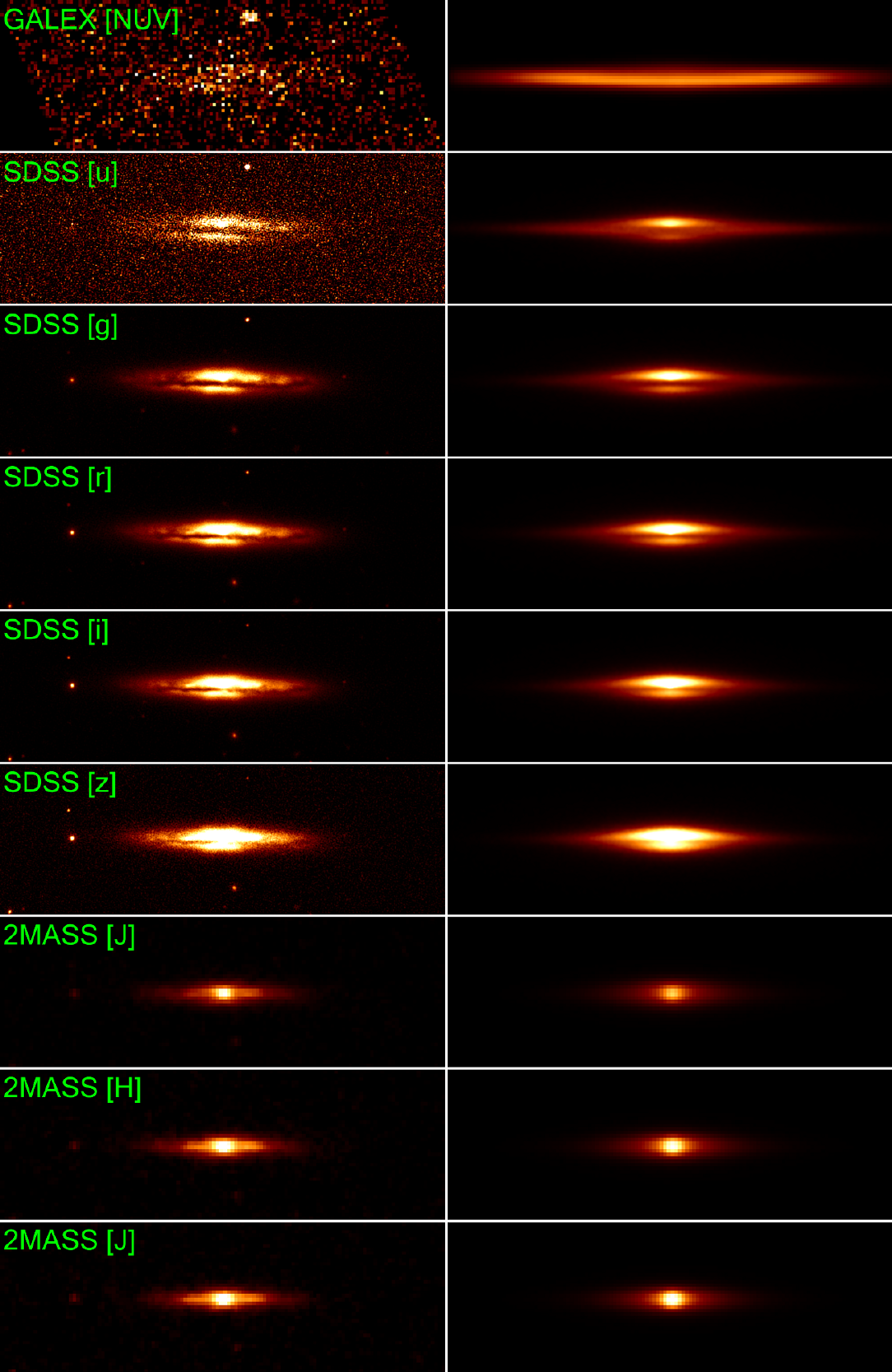}}\hfill
\subfigure{\includegraphics[width=0.46\textwidth]{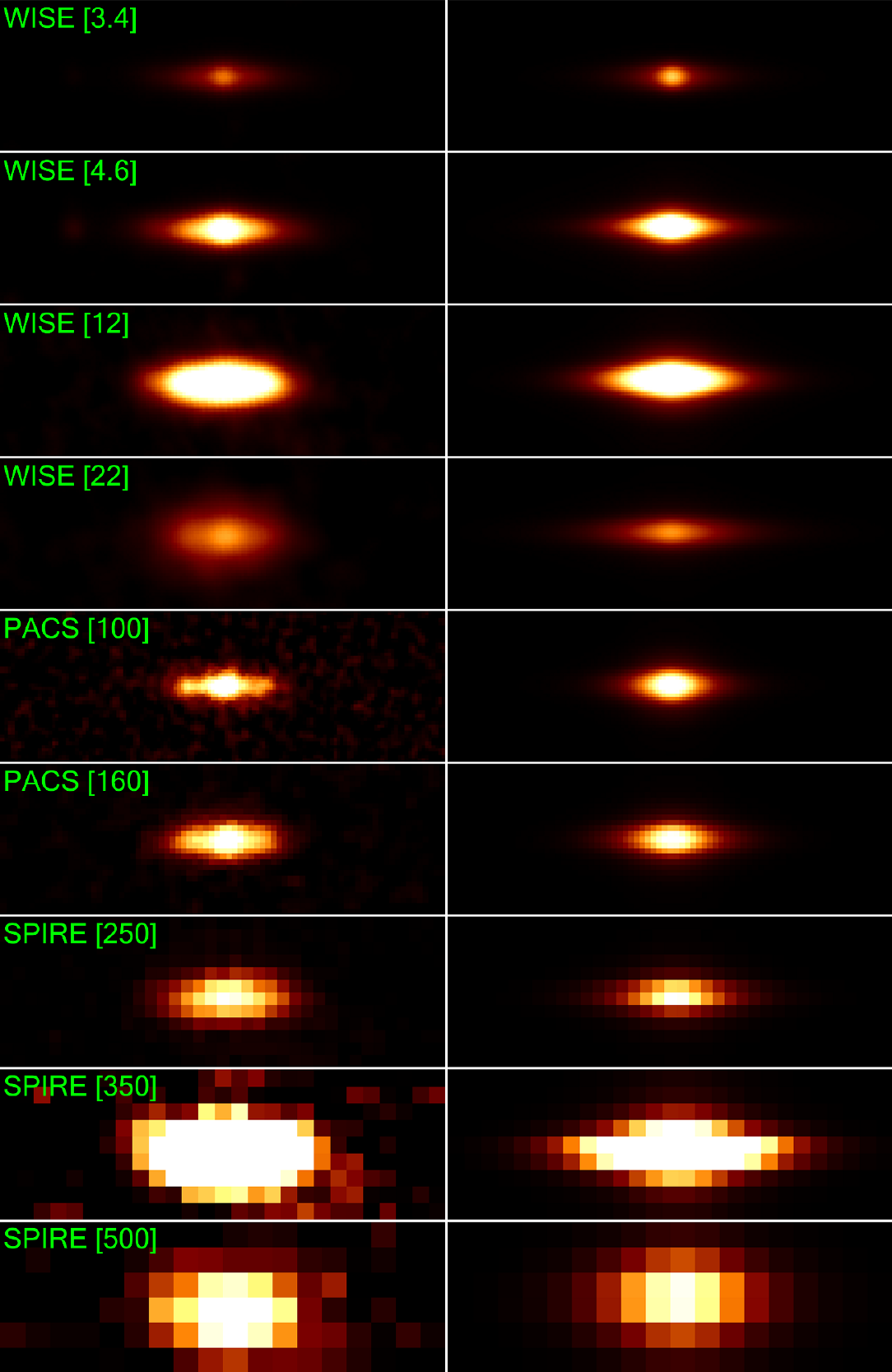}}
\hspace*{\fill}
\caption{Same as Figure~{\ref{comparison_IC4225.fig}}, but now for NGC\,5166.}
\label{comparison_NGC5166.fig}
\end{figure*}

In Figures~{\ref{comparison_IC4225.fig}} and \ref{comparison_NGC5166.fig} we compare a set of  images corresponding to the Case~B models to the actual observed images of the two galaxies in different wavebands across the entire wavelength range calculated in the simulations. The simulated images have been convolved with the appropriate point spread function: for GALEX, SDSS, 2MASS and WISE we used a simple Gaussian with the correct FMHM, for the PACS and SPIRE data we used the kernels provided by \citet{2011PASP..123.1218A}.

From the GALEX NUV image it is clear that for IC\,4225, the unobscured young stellar component is considerably more  concentrated in the center, than what is assumed in our model. Additionally, the MIR and FIR images seem to suggest that, especially the warm dust, is also concentrated closer to the centre of the galaxy. It is clear that the dust disk as obtained in \citetalias{2014MNRAS.441..869D}, is clearly not a realistic model for this galaxy. While it successfully reproduces the optical images, the disk is too extended to reproduce the UV and MIR/FIR observations. Interestingly, our radiative transfer model was capable of recovering the $500\ \mu$m flux density. However, it seems that the model does not really predict the correct morphology: the model image at 500~$\mu$m is spatially extended, while the observed SPIRE 500~$\mu$m image seems much more centrally concentrated.

For NGC\,5166 the comparison between model and observations is much more satisfactory. In general, the geometry of the model images seems to match the observations reasonably well. While less obvious due to the low signal-to-noise image, the NUV images seem to be in better agreement than was the case for IC\,4225. Apart from the boxy/peanut shaped bulge and some small bright, additional features, probably projected spiral arms, the near-infrared images also match quite well. These off-centre features also seem to be slightly noticeable at 100 $\mu$m but, in general, the model is in good agreement with the observations.

\subsection{Case~C: UV--NIR + ModBB models}
\label{CaseC.sec}

To get some more insight into the missing component in our model we have added two additional components to the previous model. 

First, to explain the underestimated emission in the FIR/submm regime, we fit a modified black body through the residual {\it{Herschel}} fluxes and add this component to our model. As in both galaxies the $500\ \mu m$ flux was recovered, this leads to a small overestimation of the cold dust contribution in this model. From this fit we can determine the amount of cool dust that is missing and the associated dust temperature. For IC\,4225 we require an additional $9 \times 10^6~M_\odot$ at a temperature of 21~K to explain all FIR data points, which amounts to about one third of the total dust mass. A significant part of the dust of IC\,4225 should hence be distributed in such a way that it does not add significantly to the large-scale attenuation but has a significant contribution in the FIR part of the SED, such as quiescent compact clumps. {{Combining both components, the total dust mass of NGC\,5166 is $3.0\times10^7~M_\odot$, 75\% larger than the value estimated using MAGPHYS ($1.7\times10^7~M_\odot$, see Table~{\ref{Properties.tab}}). The difference is probably the result of the overabundance of cold dust in our model (due to the extended dust disk, our model overestimates the observed flux densities at the longest wavelengths).}}

In the case of NGC\,5166, the best fit returns $1.5 \times 10^7~M_{\odot}$ of dust at a temperate of 20~K. However, it should be noted that this overestimates some of the observed {\it{Herschel}} fluxes and should be considered as an upper boundary for the missing dust mass. Therefore, at most one fourth of the total dust mass of NGC\,5166 is locked up in quiescent compact clumps.  {{In this case, the total dust mass amounts to $6.3\times10^7~M_\odot$, in very good agreement with the MAGPHYS value of $7.7\times10^7~M_\odot$ (Table~{\ref{Properties.tab}}).}}

Secondly, the WISE 22 $\mu$m and IRAS 25~$\mu$m mid-infrared fluxes of both galaxies are underestimated by the model. Emission at these wavelengths is dominated by obscured star formation. Some obscured star formation is already included in the Case~B models. Indeed, the observed NUV flux was fitted by adding a young stellar component, and a significant part of the corresponding UV emission was attenuated by the diffuse dust component. This additional dust heating and corresponding mid-infrared emission is clearly visible as the difference between the Case~A and Case~B lines in Figure~{\ref{SEDs.fig}}, but not sufficient to explain the observed mid-infrared flux densities. The remaining mid-infrared excess can be modelled as completely obscured star forming regions \citep{2014A&A...571A..69D}. We could opt to add another modified black body, but we preferred to use an ultra-compact H{\sc{ii}}-region template from \cite{2006A&A...458..405G}. Since this component is only constrained by the 22 and 25~$\mu$m fluxes, the corresponding dust masses are poorly constrained (in any case, they are negligible compared to the cool dust mass budget).
 
The Case~C models for both galaxies are represented as the {{dotted}} lines in Figure~{\ref{SEDs.fig}}. We would like to stress that this model is an {\it{ad hoc}} model that holds no ``predictive'' power. This is unlike previous models, where the entire SED is essentially derived in a self-consistent way from NUV/optical information only.

\section{Discussion and conclusions}
\label{disc.sec}

The ambition of this paper was to use the state-of-the-art, both observationally and in radiative transfer modelling, to investigate the so-called dust energy balance problem for edge-on spiral galaxies. For both galaxies, IC\,4225 and NGC\,5166, the observed SED shows a clear FIR excess compared to a standard model that is based only on optical images, and hence only contains evolved stellar populations. This is similar as what is found by various teams \citep[e.g.,][]{2000A&A...362..138P, 2001A&A...372..775M, 2004A&A...425..109A, 2005A&A...437..447D, 2008A&A...490..461B, 2010A&A...518L..39B, 2011A&A...527A.109P, 2012A&A...541L...5H}. However, things change when a young stellar population is included in the modelling: while a substantial FIR excess still remains for IC\,4225, the FIR fluxes are recovered almost exactly for our model of NGC\,5166. For the latter galaxy, the Case~B model does not only reproduce the observed SED, the predicted images over the entire UV--submm wavelength range also provide a good match to the observed images. For both galaxies, the SED can be forced to fit the data by adding an obscured star formation template and a component of cool dust that dust not affect the large-scale extinction. The latter could be interpreted as a population of dense and quiescent clumps.

Why these two galaxies show such a different behaviour is an interesting question. It is possible that this is a consequence of the modelling. For both galaxies, we used exactly the same approach, adopted the same assumptions and used the same data as input. Moreover, the FitSKIRT fitting of radiative transfer models to the optical images in both cases resulted in fits of similar (and satisfactory) quality. One aspect that is different, though, is the different angular size of the galaxies. 

IC\,4225 is the smallest galaxy, in angular size, of the sample studied in \citetalias{2014MNRAS.441..869D}. As a result, some of the parameters in the model were not constrained as well as for most of the other galaxies. In particular, the derived intrinsic properties of the dust disk in IC\,4225 are rather exceptional: the dust disk scale length retrieved is approximately 10~kpc, which is especially large when compared to the more moderate stellar disk with a scale length of 3.4~kpc. However, when using the uncertainties on the derived parameters from optical modelling, the model with the highest FIR output is just within range of the observed values.

NGC\,5166, on the other hand, is one of the largest galaxies in the sample, and the recovered star--dust geometry is more in line with the average results of other studies of edge-on spiral galaxies \citep{1999A&A...344..868X, 2007A&A...471..765B}. While this difference in apparent size and hence resolution might be partly responsible for the different results, it remains strange that NGC\,5166 shows hardly any FIR/submm excess. Indeed, this is the galaxy with the most ``regular'' star--dust geometry, and one would expect that this galaxy would show a FIR/submm excess as most other ``regular'' edge-on spiral galaxies. The main difference between our analysis, compared to previous work on other edge-on galaxies, is a more self-consistent radiative transfer modelling in the optical, since we fit our models to multiple bands simultaneously. Whether or not this is an important aspect remains to be investigated.

While we have modelled both galaxies in the same way, it is not impossible that these two systems are intrinsically different. For example, the compactness of the UV and FIR emission in both galaxies is fairly different: for IC\,4225 the emission is considerably more centrally concentrated than for NGC\,5166. Also the spatial distribution of the stars is not the same: the S\'ersic indices of the bulges are noticeably different, at $1.1\pm0.1$ and $5.8\pm0.9$ for IC\,4225 and NGC\,5166 respectively. This could point to {{different evolutionary mechanisms}} of both galaxies that could result in a different ISM structure (i.e., with a different degree of clumpiness and substructure), and a difference in the ratio of obscured versus unobscured star formation.

The main conclusion of our investigation is that, instead of gaining more insight in the driving mechanism for the dust energy balance problem, we seem to find that the problem might be more complex than anticipated or hoped for. If we want to get a more detailed insight in the mechanisms involved in the dust energy balance, it is necessary to perform this panchromatic radiative transfer modelling on a larger scale. Up to now, such modelling has almost exclusively been applied to individual galaxies, with different models, assumptions and codes making the comparison or search for correlations essentially impossible. A uniform modelling effort on a larger sample of edge-on galaxies would be highly desirable. 

One necessary tool for such an effort is the availability of radiative transfer fitting codes that can efficiently search large parameter spaces and work semi-automatically; the work by \citetalias{2014MNRAS.441..869D} demonstrates that this is nowadays possible. Another crucial ingredient is obviously the availability of suitable targets. Excellent catalogs of edge-on spiral galaxies are available nowadays \citep{2006A&A...445..765K, 2014ApJ...787...24B}, and the fact that relatively large areas of the extragalactic sky have been observed with {\it{Herschel}} in the frame of projects such as H-ATLAS opens possibilities.

The combined HEROES \citep{2013A&A...556A..54V} and NHEMESES \citep{2012IAUS..284..128H} samples contain 19 nearby edge-on spiral galaxies that have been observed with PACS and SPIRE. These galaxies are the ideal candidates for a follow-up study as they have a spatial resolution that is an order of magnitude better than the two galaxies studied in this paper. We are currently collecting the necessary optical data to start the modelling of this sample. 

Apart from systematically modelling larger sets of edge-on spiral galaxies, there are other options that need further exploration. One interesting option is looking at face-on spiral galaxies \citep{2014A&A...571A..69D}. In this case, smaller scale structures remain present in the observations instead of getting smoothed out over the line of sight. The disadvantage of this approach is that there is no direct analogue for the dust lanes in edge-on spiral galaxies, by which the dust mass can be estimated. Additionally, because of the lack of information on the vertical distribution, assumptions have to be made. These can be extracted from systematic studies of edge-on spiral galaxies \citep{1999A&A...344..868X, 2007A&A...471..765B, 2014MNRAS.441..869D, 2014ApJ...787...24B}, with additional priors or constraints provided by occulting galaxy studies \citep{1992Natur.359..655W, 2009AJ....137.3000H, 2013MNRAS.433...47H, 2013A&A...556A..42H} or power spectrum analyses \citep{2010ApJ...718L...1B, 2012A&A...539A..67C}.

Another necessary follow up study is the influence of the assumed dust model. While it is clear that changing the dust opacities and grain composition might not change the difference between both galaxies extremely, the overall influence of the made assumptions on the global SED was not investigated in this paper. 

Finally, a different path that needs further inspection is modelling of mock observations using radiative transfer simulated galaxies. The use of artificial data, in which the intrinsic properties are known and compared to those obtained after the radiative transfer modelling, can yield interesting clues on biases or systematic effects in the modelling. The pioneering work by \citet{2000A&A...353..117M} and \citet{2002A&A...384..866M}, who investigated the effects spiral arms and a clumpy ISM respectively, was still based on relatively uncomplicated toy models. 

A more advanced option is using the results of 3D hydrodynamical simulations, which have managed to produce more realistic spiral galaxies in the past few years \citep{2011ApJ...742...76G, 2013MNRAS.436.1836R, 2014MNRAS.437.1750M, 2015MNRAS.446..521S}. Several teams have started using such hydrodynamical simulations to investigate the systematic effects related to radiative transfer effects \citep[e.g.,][]{2015MNRAS.446.1512H}. Our initial results based on applying FitSKIRT to two of these models seems to suggest that the complex and inhomogeneous dusty medium in spiral galaxies could ``hide'' significant amounts of dust \citep{2015A&A...576A..31S}. One aspect, very relevant for the present study, that could be also investigated using such models is the role of resolution in radiative transfer modelling. 

Given the various strengths and weaknesses of each these approaches, pursuing all three options might be the best approach to gain a conclusive answer on the driving mechanisms causing the dust energy balance.

\section*{Acknowledgments}
\label{ack.sec}
G.D.G., M.B., I.D.L.\ and S.V.\ gratefully acknowledge the support of the Flemish Fund for Scientific Research (FWO-Vlaanderen). M.B., J.F.\ and T.H.\ acknowledge financial support from the Belgian Science Policy Office (BELSPO) through the PRODEX project ``Herschel-PACS Guaranteed Time and Open Time Programs: Science Exploitation'' (C90370). This work has been realised in the frame of the CHARM framework (Contemporary physical challenges in Heliospheric and AstRophysical Models), a phase VII Interuniversity Attraction Pole (IAP) programme organised by BELSPO, the BELgian federal Science Policy Office. We thank Catherine Vlahakis and Simone Bianchi for comments and suggestions. {{The anonymous referee is gratefully acknowledged for a detailed referee report that improved the content and presentation of this work.}}

The {\it{Herschel}}-ATLAS is a project with Herschel, which is an ESA space observatory with science instruments provided by European-led Principal Investigator consortia and with important participation from NASA. The H-ATLAS website is \url{http://www.h-atlas.org}.

PACS has been developed by a consortium of institutes led by MPE (Germany) and including UVIE (Austria); KU Leuven, CSL, IMEC (Belgium); CEA, LAM (France); MPIA (Germany); INAF-IFSI/OAA/OAP/OAT, LENS, SISSA (Italy); IAC (Spain). This development has been supported by the funding agencies BMVIT (Austria), ESA-PRODEX (Belgium), CEA/CNES (France), DLR (Germany), ASI/INAF (Italy), and CICYT/MCYT (Spain).

SPIRE has been developed by a consortium of institutes led by Cardiff University (UK) and including Univ. Lethbridge (Canada); NAOC (China); CEA, LAM (France); IFSI, Univ. Padua (Italy); IAC (Spain); Stockholm Observatory (Sweden); Imperial College London, RAL, UCL-MSSL, UKATC, Univ. Sussex (UK); and Caltech, JPL, NHSC, Univ. Colorado (USA). This development has been supported by national funding agencies: CSA (Canada); NAOC (China); CEA, CNES, CNRS (France); ASI (Italy); MCINN (Spain); SNSB (Sweden); STFC, UKSA (UK); and NASA (USA).

Funding for the SDSS and SDSS-II has been provided by the Alfred P. Sloan Foundation, the Participating Institutions, the National Science Foundation, the US Department of Energy, the National Aeronautics and Space Administration, the Japanese Monbukagakusho, the Max Planck Society, and the Higher Education Funding Council for England. The SDSS Web Site is \url{www.sdss.org}. The SDSS is managed by the Astrophysical Research Consortium for the Participating Institutions. The Participating Institutions are the American Museum of Natural History, Astrophysical Institute Potsdam, University of Basel, University of Cambridge, Case Western Reserve University, University of Chicago, Drexel University, Fermilab, the Institute for Advanced Study, the Japan Participation Group, Johns Hopkins University, the Joint Institute for Nuclear Astrophysics, the Kavli Institute for Particle Astrophysics and Cosmology, the Korean Scientist Group, the Chinese Academy of Sciences (LAMOST), Los Alamos National Laboratory, the Max-Planck- Institute for Astronomy (MPIA), the Max-Planck-Institute for Astrophysics (MPA), New Mexico State University, Ohio State University, University of Pittsburgh, University of Portsmouth, Princeton University, the United States Naval Observatory, and the University of Washington.

This research has made use of SAOImage DS9, developed by Smithsonian Astrophysical Observatory.

\bibliography{H-atlas}

\end{document}